# Electron transport measurements in liquid xenon with Xenoscope, a large-scale DARWIN demonstrator

L. Baudis[1], Y. Biondi[1,2,a], A. Bismark[1], A. P. Cimental Chávez[1], J. J. Cuenca-García[1], J. Franchi[1], M. Galloway[1], F. Girard[1,b], R. Peres[1,c], D. Ramírez García[1], P. Sanchez-Lucas[1,3], K. Thieme[1,4], C. Wittweg[1]

[1] Department of Physics, University of Zurich, Winterthurerstrasse 190, 8057 Zurich, Switzerland
[2] Karlsruhe Institute of Technology, Karlsruhe, Germany
[3] University of Granada, Granada, Spain
[4] University of Hawai'i at Mānoa, Honolulu, USA

**Abstract** The DARWIN observatory is a proposed next-generation experiment with 40 tonnes of liquid xenon as an active target in a time projection chamber. To study challenges related to the construction and operation of a multi-tonne scale detector, we have designed and constructed a vertical, full-scale demonstrator for the DARWIN experiment at the University of Zurich. Here, we present the first results from a several-months run with 343 kg of xenon and electron drift lifetime and transport measurements with a 53 cm tall purity monitor immersed in the cryogenic liquid. After 88 days of continuous purification, the electron lifetime reached a value of $(664 \pm 23)\,\mu$s. We measured the drift velocity of electrons for electric fields in the range (25–75) V/cm, and found values consistent with previous measurements. We also calculated the longitudinal diffusion constant of the electron cloud in the same field range, and compared with previous data, as well as with predictions from an empirical model.

## 1 Introduction

The next-generation liquid xenon (LXe) detector DARWIN [1,2] will operate a dual-phase time projection chamber (TPC) filled with 40 tonnes of xenon, to probe the parameter space for weakly interacting massive particle dark matter down to the neutrino fog [2–4]. With the large target mass and projected low background level, competitive searches for neutrinoless double-beta decay of $^{136}$Xe [5,6], solar axions, bosonic dark matter and dark photons are viable. Solar $^8$B and supernova neutrinos can be detected via coherent neutrino-nucleus scattering [7,8], while solar *pp*-neutrinos can be observed with high statistics at low energies via their scattering off atomic electrons [9].

At its core, DARWIN will operate a cylindrical dual-phase TPC, almost entirely filled with LXe, with a thin gaseous xenon (GXe) layer at the top. Energy depositions generate scintillation light and ionization electrons whose signals, denoted as S1 and S2, respectively, are measured with light sensors covering the top and bottom faces of the cylindrical detector. The dual signal readout enables energy and position reconstruction, as well as background discrimination. A $\mathcal{O}(100)$ V/cm electric drift field is applied throughout the LXe volume between a cathode and a gate electrode, suppressing electron recombination while moving them towards the gate below the liquid–gas interface. A stronger $\mathcal{O}(10)$ kV/cm extraction field between the gate and the anode accelerates the electrons into the GXe, where they produce an electroluminescence signal proportional to the number of electrons reaching the liquid–gas interface.

Electron losses mostly occur due to attachment to impurities during the drift. A depth-dependent widening of the S2 arises from electron cloud diffusion during the drift. Longitudinal diffusion (parallel to the electron cloud propagation) affects the ability to distinguish multiple S2s occurring at different depths ($z$) within a single event. Additionally, lateral diffusion can impact position reconstruction in the *x*-*y* plane. Since such signatures are expected in background events with multiple interactions of neutrons or γ-rays, diffusion impacts the background rejection capability of a detector.

[a] e-mail: yanina.biondi@physik.uzh.ch (corresponding author)
[b] e-mail: frederic.girard@physik.uzh.ch (corresponding author)
[c] e-mail: ricardo.peres@physik.uzh.ch (corresponding author)

With its 2.6 m in drift length and 2.6 m diameter, the cylindrical DARWIN TPC will be significantly larger than currently operative LZ [10], PandaX-4T [11] and XENONnT [12], with dimensions up to 1.5 m drift and diameter. To study several challenges related to the increased dimensions, such as high-voltage delivery, electron loss, diffusion of the electron cloud and light attenuation over large distances, the Xenoscope facility was designed and built as a vertical scale technical demonstrator [13]. Section 2 presents the Xenoscope facility during its first science phase. The facility was equipped with a 53 cm tall purity monitor, described in Sect. 3, which was targeted at measuring electron drift and quantifying the xenon purification capabilities along with the electron transport measurements. Section 4 discusses the results from the electron drift lifetime measurement for different purification regimes and presents the fit of an electron drift lifetime model to the data. It also reports on measurements of the electron drift velocity and of the longitudinal diffusion coefficient of electron clouds at different drift fields. Conclusions and an outlook are given in Sect. 5.

## 2 The xenoscope facility

Xenoscope can house up to 400 kg of LXe in a double-walled stainless steel cryostat. The facility, its subsystems, and the outcome from the first commissioning run are described in Ref. [13]. Xenoscope was first equipped with a purity monitor (Sect. 3) fully submerged in LXe, while the cryostat aspect ratio was chosen to allow for the operation, in the next phase of the project, of a 2.6 m two-phase TPC, with the primary goal of demonstrating the drift of electrons in LXe over this distance for the first time. A computer-aided design (CAD) rendering of the cryostat with the purity monitor is shown in Fig. 1.

The facility includes a gas purification system with a series of filters and a commercial zirconium alloy getter. The LXe is extracted at the top of the liquid column, where the impurity concentration is higher. It is evaporated in the heat exchanger system and circulated through the purification system at a fixed flow. The purified xenon is recondensed in the heat exchanger and reintroduced in the cooling tower, which comprises a pulse tube refrigerator (PTR) connected to a cold head mounted atop the cooling chamber. The xenon is then directed to the bottom of the cryostat. A slow control system built from open-source software oversees and sends alarms on relevant parameters.

Two system upgrades were performed prior to the installation of the purity monitor. First, a pre-cooler was manufactured and installed at the top of the inner cryostat vessel to provide additional peak cooling power, and thus reduce the system cooldown and xenon liquefaction time during filling by a factor of 4.25. The design of the pre-cooler and details

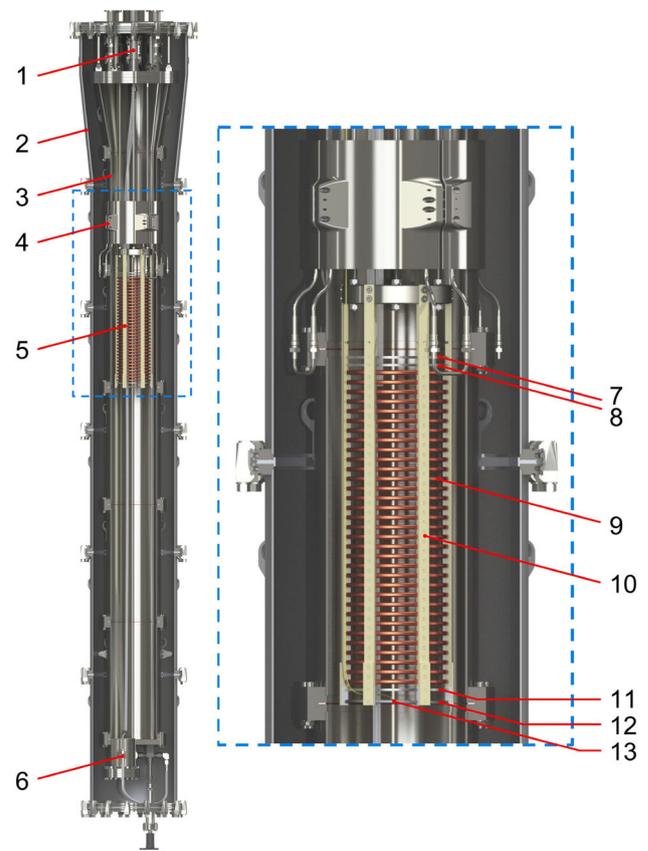

**Fig. 1** The purity monitor in the Xenoscope cryostat. Legend: (1) top flange; (2) outer vessel; (3) inner vessel; (4) pre-cooler; (5) purity monitor; (6) BoX recuperation line; (7) anode; (8) anode grid; (9) field-shaping rings and resistor chain; (10) support pillars; (11) cathode grid; (12) cathode disk; (13) photocathode and optical fibre

of its commissioning are presented in Appendix A. Furthermore, a gravity-assisted recuperation and storage system for LXe, Ball of Xenon (BoX), was deployed to allow for the storage of up to 450 kg of xenon at room temperature, as well as for recuperation in liquid phase. The latter enhances the speed of the recuperation process by a factor $\sim 8$ compared to gaseous recuperation to a bottle array via cryogenic pumping. More details of its design and performance are presented in Appendix B.

## 3 The purity monitor and measurements

Common impurities in commercially available xenon consist of parts-per-million (ppm) levels of $O_2$, $N_2$, $H_2O$, CO, as well as organic molecules [14]. Additionally, detector and subsystem materials introduce impurities by outgassing. The purification of xenon prevents electron losses via their attachment to electronegative impurities and allows to achieve high light and charge yields.

Most purity monitors measure the charge deficit of an initially known population of electrons after their drift through the liquid. By comparing the number of electrons before, $N_0$, and after the drift, $N(t_d)$, an indirect measurement of the impurity concentration in LXe can be achieved. The deficit can be modelled as a decaying exponential:

$$N(t_d) = N_0\, e^{-t_d/\tau}, \tag{1}$$

where $\tau$ is the electron drift lifetime. It relates to the concentration of electronegative impurities as:

$$\tau = \frac{1}{\sum_i k_i n_i}, \tag{2}$$

where $k_i$ is the attachment rate specific to the impurity type in units of volume per mol per time, usually given in L/(mol s), $n_i$ is the impurity concentration given in mol/L, and the sum extends over the different electronegative species $i$ in the LXe. The attachment rate coefficient depends on the electric field strength.

A schematic of the working principle of the purity monitor is shown in Fig. 2, left. An optical fibre transmits the light from a xenon flash lamp to the centre of a photocathode. The incident photons produce electrons via the photoelectric effect. The electrons are drifted via extraction, drift, and collection electric fields, generated by four biased electrodes. The first drift region (1) is located between the cathode (with the photocathode in the centre) and the cathode grid; the second region (2) extends up to the anode grid; the third region (3) extends from the anode grid to the anode. The charges induce a current signal in the cathode as they drift towards the cathode grid. The screening grids prevent current induction in the cathode and anode when the electrons are drifting along the second region. Once the electrons reach the third drift region, a second signal is generated at the anode, until the electrons are fully collected. Two electronic circuits amplify and convert the induced currents to voltage signals.

The data is acquired and digitised, triggered by the pulse generator which also starts the discharge in the xenon flash lamp, with a window of 100 μs for the anode and cathode waveforms. Once digitised, the voltage signals are integrated to obtain the charges, i.e. the number of extracted and surviving electrons. With the induced charges and the time between the two signals, which corresponds to the drift time for the applied electric field, the electron drift lifetime can be inferred by solving numerically the equation:

$$\frac{Q_A}{Q_C} = \frac{t_1}{t_3} e^{-(t_1+t_2+t_3)/\tau} \frac{(e^{t_3/\tau}-1)}{(e^{-t_1/\tau}-1)}. \tag{3}$$

Here, $Q_A$ and $Q_C$ are the charges from the integrated signals measured in the anode and cathode, respectively, $t_1$ is the rise time of the first signal, $t_2$ is the time between the minimum of the signal in the cathode and the rise time of the signal

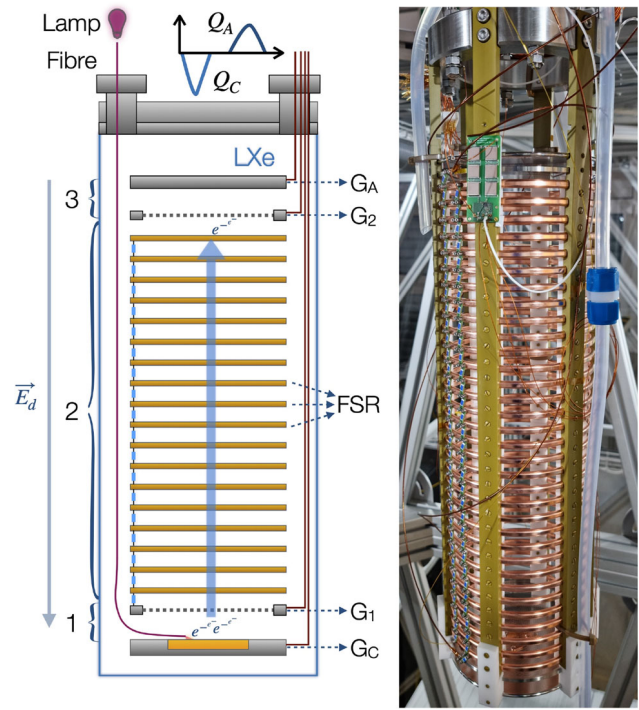

**Fig. 2** (Left): Schematic of the purity monitor. A pulse generator triggers a flash from the xenon lamp and the light is transmitted through an optical fibre to the photocathode, where photoelectrons are produced. The electrons are extracted, transported and collected by three electric fields, defined by the cathode ($G_C$) and cathode grid ($G_1$), the anode grid ($G_2$) and the anode ($G_A$). In the longest region (2), the field shaping rings (FSR) maintain the uniformity of the drift field $\mathbf{E_d}$ in the vertical direction. (Right): Assembled purity monitor in Xenoscope

in the anode, with $t_3$ the time from $t_2$ up to the maximum of the anode signal. Given the motion of the charges, the signal in the cathode has negative polarity, while in the anode the polarity is positive. Figure 3 shows an example of signals acquired in LXe from the cathode and anode at 40 slpm along with the three drift times.

The design of the Xenoscope purity monitor is described in detail in Refs. [13,15], and the assembled module is shown in Fig. 2, right. It features a field cage built with high conductivity, oxygen-free copper rings, supported by six polyamide-imide pillars. The rings are connected by a resistor chain of 5 GΩ impedance each, and enclose a cylindrical drift region of 15 cm ⌀ × 53.1 cm. The cathode and anode grids consist of hexagonally-patterned, etched stainless steel meshes with high optical transparency (∼ 93%), while the cathode and anode are solid stainless steel disks.

### 3.1 Optical components and photocathode

The utilised lamp is a 60 W xenon flash lamp with a built-in reflective mirror (model number L7685) from *Hamamatsu* [16]. The window is a single sapphire crystal allowing

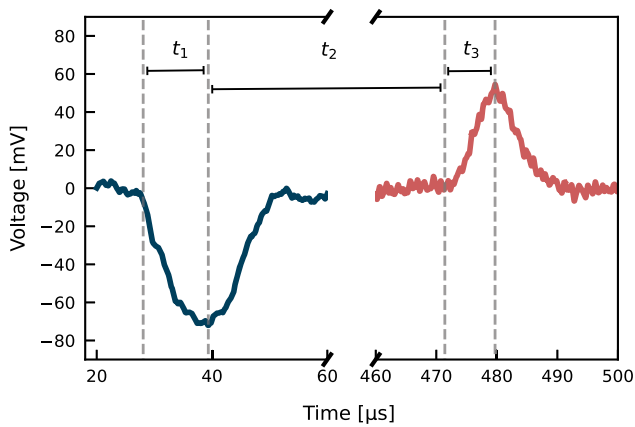

**Fig. 3** Signals acquired at 40 slpm xenon recirculation speed. The rise time of the cathode signal (blue) is taken as $t_1$, and the time interval between the minimum of the cathode signal and the start of the anode signal (red) is taken as $t_2$, with half of the charge cloud completely collected at $t_3$. These values are later used to calculate the electron drift lifetime

short wavelength light (∼ 190 nm) to reach the output of the lamp, with a spectral emission from 190 to 2000 nm. The lamp generates a discharge which excites the gas producing scintillation, with reflective mirrors directing photons from all directions towards the output. The xenon lamp can be triggered internally, or externally via a pulse generator. The intensity of the light emission is adjusted by setting the voltage for the discharge in the lamp between 600 V and 1 kV. The selection of the latter maximises the number of produced electrons.

Measures were adopted to mitigate the electronic noise produced in the signal waveforms by the external trigger: the xenon flash lamp was rehoused in a stray electromagnetic interference box, and galvanic insulation and ferrite filters were added to the trigger line. The lamp was placed outside the cryostat, with an optical fibre carrying the light from its output to the surface of the photocathode. A UV grade sapphire lens produced by *Hamamatsu* was placed at the output of the lamp to collimate the light to the optical fibre. The selected fibre was 600 μm in diameter, ultra-high vacuum rated with a polyimide buffer from *LewVac* [17], and solarisation resistant. One of the critical parts of a purity monitor is the photocathode, as it directly impacts the size of the initial signal. It consists of a thin layer of a low work function metal, deposited on a quartz substrate that has low absorption of UV photons [18]. The photocathode was produced in-house using a turbomolecular pumped coater Q150T Plus from *Quorum* [19]. The desired thickness of the layer was monitored with a quartz crystal microbalance. The coater was used to produce gold photocathodes of 50 nm thickness on a 2 mm thick quartz substrate, with a diameter of 30.00(5) mm. The deposition of a 5 nm thick layer of titanium was required for adhesion to the substrate. The choice of thickness was based on the effective probe depth of gold layers, and previous works [20]. Additional technical details can be found in [15,21].

The photocathodes were tested in a vacuum setup, where the xenon lamp was flashed onto the photocathode material and the induced current was measured. The photocathodes showed high yields, with an increasing quantum efficiency with time when exposed to light, which did not revert back in subsequent tests. The increase in quantum efficiency of the photocathode with UV-light exposure was also observed in Ref. [20].

### 3.2 Current readout and signal processing

The readout electronics amplify the induced currents from the cathode and anode and are placed inside the cryostat to avoid signal losses along the 9 m signal cables. The circuits were designed together with the Electronics Workshop at the University of Zurich. The circuit consists of an AC-coupling component, a transimpedance amplifier, and a final voltage amplifier with a 50 Ω impedance termination to match the one from the data acquisition. The transimpedance and voltage amplifiers are implemented with two low-cost operational amplifiers, model AD8066 from *Analog Devices* [22]. An AC-coupling filter in the circuit board removes high-frequency noise, which enhances the signal quality, and the AC-coupling removes the DC component of the HV applied to the electrodes. The usage of a transimpedance amplifier, in contrast to a charge amplifier, allows for more precise timing and signal spread analyses due to its small resistive-capacitive constant (RC) and short rise time of 0.14 ms. However, due to its fast response, a low-pass filter for frequencies below 800 kHz was applied to the signals to decrease the electronic noise induced by, e.g., the pulse generator that triggers the lamp, two temperature sensors, and the uninterruptible power supply. The preamplifier operates in current mode, as the capacitance discharges rapidly, resulting in an output voltage proportional to the instantaneous current. The frequency response of the readout electronics was benchmarked, with a negligible effect on the signal shape due to the 100 MHz bandwidth.

The performance of the readout electronics was tested in a climate chamber in steps of 10 K from room temperature down to 190 K. The calibration showed a charge amplification of 0.18 fC/(mV μs), with good thermal stability. Additionally, the RC decay constant of the circuit, which could be a source of systematic error for time measurements, was estimated at ∼ 150 ns by feeding a 2 μs wide square pulse to the circuit.

An oscilloscope, *Teledyne LeCroy* model Waverunner 6104A [23], and an analog-to-digital converter from *CAEN*, model v1724 [24], acquired the waveforms produced by the cathode and anode readout. Each acquisition consisted of the average of 1000 waveforms acquired over 16.7 minutes to minimise the baseline noise. The signals were then processed by fitting the expected signal shape with a Gaussian distribution. In some waveforms, a noise introduced by external electronic devices could be discerned as part of the background noise, and the fit included a sine function to account for this effect, with an inferred model uncertainty of 5% for the ones where the sine fit to the noise did not converge. The current-equivalent voltage signals in the cathode and anode were integrated to obtain a charge-proportional value. The residuals of the fits were used as weights for the charge values obtained in the averaged data shown in the next section. The uncertainties in charges and times obtained in the fits were propagated to obtain the uncertainty of the electron drift lifetime value. An example of the raw anode signal at 53 V/cm drift field in region 2 is shown in Fig. 4, together with the post-processing signal with a low-pass filter. The calculated baseline and Gaussian fit of the signals are also shown.

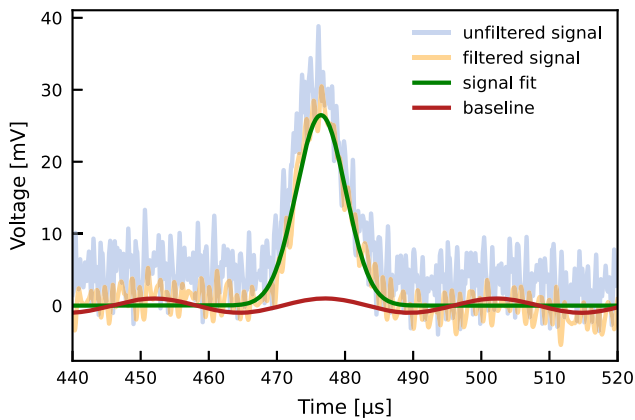

**Fig. 4** Anode signal at 53 V/cm prior to (blue) and after (orange) the low-pass filter. The calculated baseline (red) and a Gaussian fit of the signal (green) are also shown. The signal is an average over 1000 recorded waveforms

**Table 1** Electric fields, distances and times $t_i$ measured for the three regions in the PM, with voltages $-2710$ V, $-2650$ V, 0 V, and 500 V for the cathode, cathode grid, anode grid and anode, respectively, for a purification speed of 40 slpm

| Drift region $i$ | Distance [mm] | Field [V/cm] | $t_i$ [µs] |
|---|---|---|---|
| 1 | $18 \pm 1$ | $33 \pm 1$ | $12.8 \pm 0.8$ |
| 2 | $503 \pm 5$ | $53 \pm 1$ | $433.5 \pm 0.7$ |
| 3 | $10 \pm 1$ | $500 \pm 5$ | $7.6 \pm 0.7$ |

### 3.3 Measurements

Once installed in the cryostat, the purity monitor was first operated in vacuum ($\sim 1 \cdot 10^{-5}$ mbar). Data was acquired to investigate the signal shape and response in this configuration with negligible charge losses due to residual gas. The measurement additionally provided the delay time of the electronics chain, from the pulse generator for the xenon lamp to the signal amplification and readout of 18 µs.

After the calibration of the purity monitor in vacuum, gaseous xenon was flushed inside the detector and purified through recirculation in the gas system. The LXe run started with the filling of 343 kg of xenon. As the xenon recirculates through the getter, electronegative impurities are removed, and the electron drift lifetime is expected to increase in two steps: an initial exponentially increasing phase where the bulk impurities are rapidly removed, and a second phase where the change is dominated by the outgassing of materials and where the electron drift lifetime slowly increases over time. At different recirculation speeds, the electron drift lifetime reaches increasingly higher values in the second phase.

The recirculation speed was set with flows of 30 standard litres per minute (slpm), 35 slpm and 40 slpm, with the xenon lamp illuminating the photocathode with a frequency of 1 Hz. In the cryostat, the temperature and pressure were maintained around $(177.6 \pm 0.1)$ K and $(2.05 \pm 0.01)$ bar, respectively. Following the commissioning run, the displacement of the GXe compressor was reduced to increase its lifespan. This constrained the maximum purification speed to 40 slpm, compared to the 80 slpm reported in Ref. [13]. The initial impurity level in the xenon gas impacts the number of days before a signal can be seen in the purity monitor, and the first waveforms in the cathode and anode were observed after $\sim 26.5$ days.

During data taking, the cathode and cathode grid were biased at $-2710$ V and $-2650$ V, respectively. The anode grid was kept at ground while the anode was biased at 500 V. The values were selected based on COMSOL [25] simulations which yielded nearly 100% extraction efficiency of the electrons produced in the centre of the photocathode. Table 1 shows the summary of the distances, times electric fields for the extraction (1), drift (2) and collection (3) regions.

Figure 5 shows the anode and cathode signals with their integral, where the integrated signals show a step-like feature after the charges move entirely to the next drift region, or are collected in the anode. The integration corresponds to the total area of the Gaussian fit. The charge measured in the cathode corresponds to $N_{e^-} \cong 10^6$ electrons extracted from the photocathode at each pulse.

## 4 Results and discussion

The electron drift lifetime measurement campaign with the purity monitor lasted a total of 88 days. The purification was performed at 30 slpm for 46.6 days, at 35 slpm for 20.0 days, and at 40 slpm for 21.2 days. After the electron drift lifetime measurements in LXe, signals for drift fields from 25

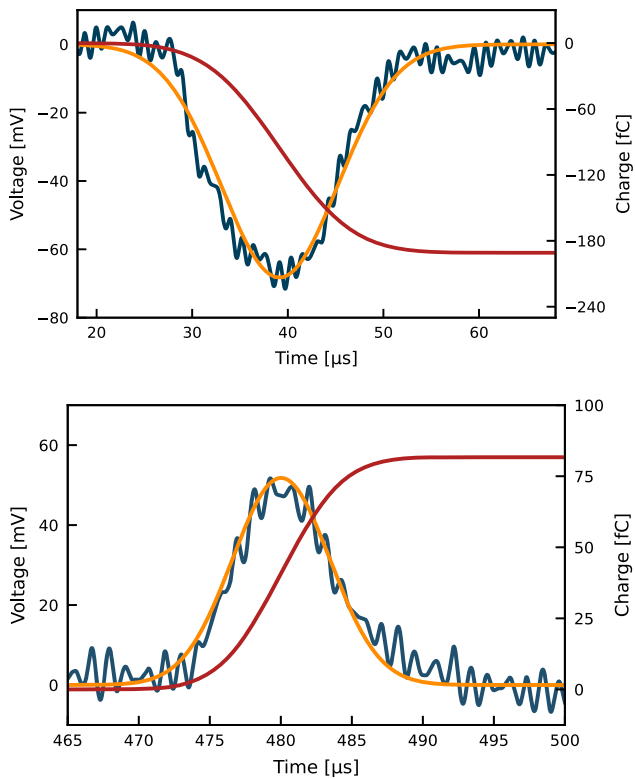

**Fig. 5** Signal readout (blue) at the cathode (top) and anode (bottom) with their respective Gaussian fits (orange) and integrated charge signals (red)

to 75 V/cm were acquired to study field-dependent electron transport properties, such as the drift velocity and longitudinal electron cloud diffusion.

### 4.1 Electron drift lifetime

Figure 6 shows the electron drift lifetime calculated with the charge signals acquired at the cathode and anode over the entire acquisition period. When the recirculation speed changes, the electron drift lifetime drops, most likely due to a change in the height of the liquid level, resulting in the release of trapped impurities in the high-surface tension region at the LXe "collar" (LXe/GXe/inner vessel interface). A drop in electron drift lifetime was also observed at 59.8 days, as expected when the GXe compressor was stopped for a period of approximately 15 m due to a communication error with the slow control software. Again, the change in liquid level most likely resulted in the sudden release of impurities from the collar. Shortly following these events, the electron drift lifetime increased exponentially to return to the outgassing-limited values.

A review of the slow control data allowed for the identification of three irregularities in the pressure and flow conditions, at 38.9 days, 52.6 days, and 80.7 days. After 88.0 days of almost continuous purification, the electron drift lifetime reached a value of $(664 \pm 23)\,\mu\text{s}$. This is consistent with the value reached in other LXe experiments, such as XENON1T [26] and LUX [27], with $660\,\mu\text{s}$ and $750\,\mu\text{s}$, respectively.

While the purity level demonstrated in Xenoscope could be sufficient to drift electrons over 2.6 m in LXe, it can be further improved by increasing the purification speed. To investigate this, a simple model of the electron drift lifetime, assuming $O_2$-like impurities, was adapted from Refs. [28,29] and fitted to the electron drift lifetime data shown in Fig. 6. Two coupled differential equations describe changes in impurity levels $M\frac{dI}{dt}$ after a time $dt$, where $M$ is the mass of xenon, and $I$ is the impurity concentration. The gas and liquid phases (denoted by the subscripts $g$ and $l$, respectively) are evaluated separately to determine the impurity concentration at time $t + dt$:

$$M_g \frac{dI_g}{dt}^{(j)} = -F_g \rho I_g + \left( \frac{\Lambda_{g,0}}{1 + \frac{t - \Delta t_g^{(j)}}{T_{1/2,g}}} + C_g \right)$$
$$+ \frac{\epsilon_1 P_C I_l}{h} - \frac{\epsilon_2 P I_g}{h}$$
$$+ M_g \Delta I_g^{(j)} \int \delta \left( t - t^{(j)} \right) dt \quad (4)$$

$$M_l \frac{dI_l}{dt}^{(j)} = -F_l \rho I_l + \left( \frac{\Lambda_{l,0}}{1 + \frac{t - \Delta t_l^{(j)}}{T_{1/2,l}}} + C_l \right)$$
$$- \frac{\epsilon_1 P_C I_l}{h} + \frac{\epsilon_2 P I_g}{h}$$
$$+ M_l \Delta I_l^{(j)} \int \delta \left( t - t^{(j)} \right) dt . \quad (5)$$

The index $j = \{0, 1, \ldots, 7\}$ indicates the regions between discontinuities, marked by the dashed lines in Fig. 6. Each equation consists of five terms. The first accounts for the purification rate, with $F$ being the purification flow, $\rho$ the density of xenon at 1 bar, 0 °C, and $I$ the concentration in impurities. The second term accounts for the time-dependent average outgassing rate from detector materials, where $\Lambda_0$ is the outgassing at time $t = 0$, $T_{1/2}$ is the decay-time of the outgassing rate and $C$ is a constant outgassing term which brings the system to an equilibrium point when $t \gg T_{1/2}$. After the sudden increase of impurity concentration in the xenon during flow changes, described by the delta function in the fifth term, the outgassing rate modelled by the second terms of the differential equations is expected to revert back to a high value, as impurities can be adsorbed in outgassed materials [29]. This is accounted for with the time-offset parameters $\Delta t^{(j)} = \sum_1^j \Delta t^{(i)}$, which are cumulative since the dataset is fitted as a whole, and $\Delta t^{(0)} = 0$. Finally, the third and fourth terms describe the exchange of impuri-ties between the gas and the liquid phases, and thus the signs are inverted between the two equations. The $\epsilon$ parameters are efficiencies of the exchange process, $P_C$ is the cooling power of the system in the absence of purification, proportional to the evaporation rate of the xenon, $P$ is the cooling power deployed by the cryogenics, proportional to the condensation rate, and $h$ is the latent heat of xenon.

The electron drift lifetime is then calculated by solving the system of differential equations for Eq. (2) and minimising simultaneously in all $j$-regions, using a least-square method. The best-fit model is displayed as a solid red curve in Fig. 6. Assuming the same initial conditions obtained from the fit of the model to the electron drift lifetime data, we obtain the purification flow-dependent electron drift lifetime predictions shown in Fig. 7.

Although the initial conditions are not likely to be reproducible in future detector runs, a number of observations can be extracted from this model. As expected, an increased purification speed would yield longer electron drift lifetimes, attained in a shorter purification time. The addition of a purification flow of the gas phase ($F_g = 2$ slpm) suggests an expected increase in electron drift lifetime of up to 15%. Therefore, prior to the start of the next phase of Xenoscope, a parallel gas extraction line inspired by the gas purification system reported in Ref. [29] was added to the gas handling system with a second flow controller, allowing for the purification of both the liquid and gas phases in the same purification loop.

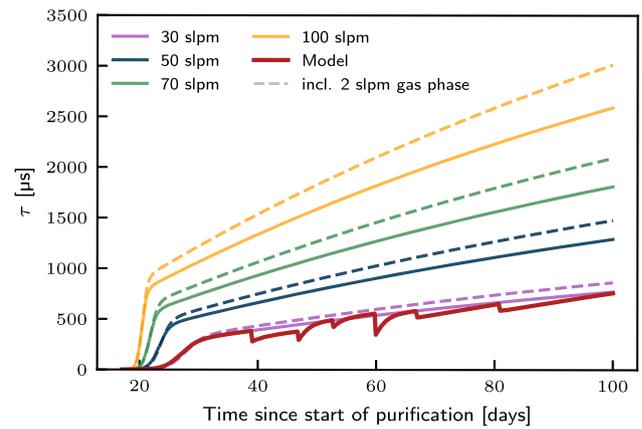

**Fig. 7** Purification flow-dependent electron drift lifetime prediction. Assuming the same operating conditions as in the measurement campaign, an increased flow (solid lines) could significantly reduce the purification time needed before the electron drift lifetime begins its exponential rise. Furthermore, the addition of a 2 slpm gas phase extraction to the purification loop can also improve the electron drift lifetime (dashed lines)

### 4.2 Electron transport

The purity monitor allows for a dedicated measurement of the arrival time of the electron cloud at the anode. The drift velocity $v_d$ of the cloud given a drift field $E_d$ is:

$$v_d = d_2/t_2. \tag{6}$$

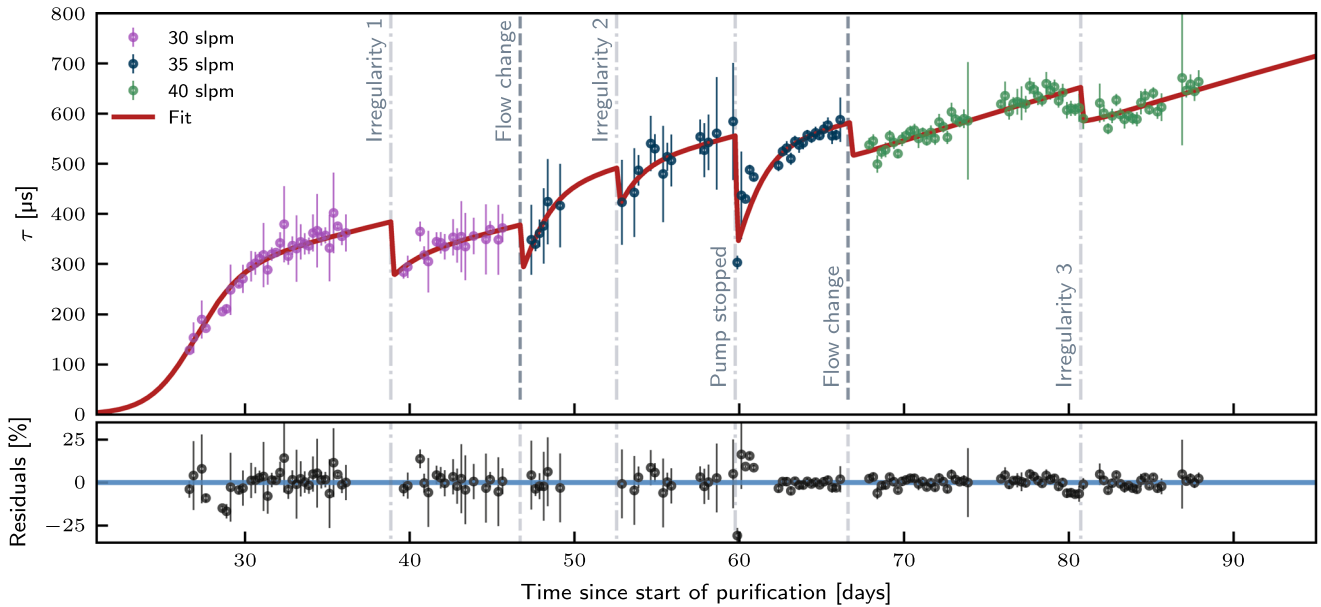

**Fig. 6** Purification flow-dependent electron drift lifetime measured in Xenoscope. The data was averaged in 6 h time bins. The dashed lines indicate a change in flow, while the dash-dotted lines indicate short-term irregularities in the pressure and flow conditions (see text). The red line shows the best-fit model from Eqs. 4 and 5, while the black points show the residuals

Considering that the accuracy of the time measurement between the extrema of the cathode and anode signals is higher, and region 3 has a fast detection, it is convenient to calculate instead:

$$v_d = (d_2 + d_3)/(t_2 + t_3). \tag{7}$$

This approach introduces an additional term, $(d_2 t_3 - d_3 t_2)/(t_2(t_2 + t_3))$, which induces a negligible 0.1% bias in the final values given the ratios between $t_3/t_2$ and $d_3/d_2$. The drift velocity can also be expressed in terms of the drift field $E_d$:

$$v_d = \mu E_d, \tag{8}$$

where $\mu$ is the electron mobility. The mobility is related to the average time, for a given temperature, density, and energy of the electrons, between elastic collisions with the xenon and electronegative impurities, and potential inelastic collisions with impurities. Thus, the acquisition of waveforms used to derive the drift velocity was performed at a constant electron drift lifetime to avoid systematic uncertainties. The acquired measurements can be compared with simple transport models for electrons: benchmark regimes can be used to visualise these dependencies for electron mobility, such as "cold electrons" and "hot electrons" [30]. In the cold electrons regime, the energies of the electrons are mostly due to the thermal bath in the xenon fluid (around 0.015 eV at 177 k [31]), and they rapidly acquire energy with increasing electric fields, resulting in a linear gain in velocity. For cold electrons, the mobility can be expressed as:

$$\mu = \frac{2}{3}\left(\frac{2}{\pi m_e k_B T}\right)^{\frac{1}{2}} e \frac{\lambda}{v}, \tag{9}$$

where $k_B$ is the Boltzmann constant, $T$ is the temperature, $e$ is the charge of the electron, $m_e$ is their mass, $v$ is the magnitude of the velocity in all directions, and $\lambda$ is the mean free path of the electrons in their collisions with atoms, inversely proportional to the number density and cross section. In contrast to cold electrons, hot electrons have gained most of their energy through acceleration by the drift field, and experience increased energy losses on their drift path due to collisions with xenon atoms, which results in a slower rate of change in their velocity. For this case, the electron mobility can be expressed as:

$$\mu = \frac{4 e \lambda}{3 v \pi^{1/2} m_e^*}, \tag{10}$$

where $m_e^*$ is the effective mass of the electrons in the medium. While these equations are not used in this work to infer properties such as electron mobility, or electron cloud diffusion,

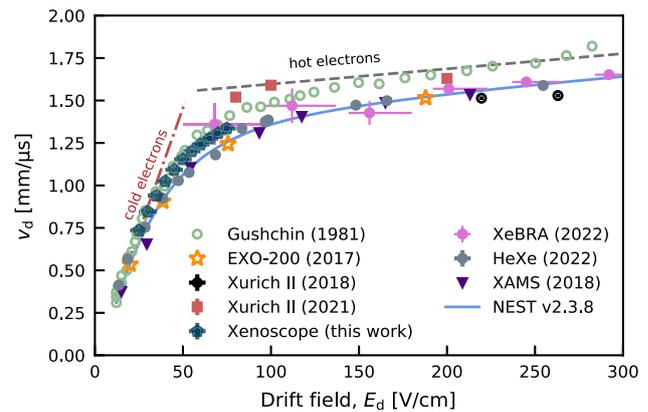

**Fig. 8** Measured drift velocity with electric field values from 25 to 75 V/cm, in steps of 5 V/cm. The results are compared to literature values from Gushchin (165 K) [32], EXO-200 (167 K) [33], Xurich II (2018, 184 K) [34], Xurich II (2021, 177 K) [35], HeXe (174 K) [36], XeBRA (173 K) [37] and XAMS (183 K). The prediction from NEST v2.3.8 [38] is shown as a solid blue curve

they are useful to scrutinise the results of our measured drift velocity and discuss potential systematic effects, discussed later in this section.

The data for this measurement was acquired at day 89, after the electron drift lifetime entered a region of slow change, and in a time interval of half an hour, to minimise systematic effects. Drift fields of 25 to 75 V/cm were scanned in steps of 5 V/cm, where the former was the threshold for a discernible signal above noise level in the cathode. The extraction field was changed to maintain the ratio between the extraction and drift field used in the electron drift lifetime data, while the collection field was fixed at 500 V/cm. Figure 8 shows the drift velocity at different fields, calculated with Eq. (7), with good agreement with previous measurements in LXe [32–37]. The prediction from NEST (Noble Element Simulation Technique) [38], based on data-driven empirical models, is also shown. The curves for cold and hot electrons are derived by using Eqs. (9) and (10), respectively, to fit the data from Ref. [32], for it covers both regimes with a high density of points.

The measured drift velocities in LXe with different setups show inconsistencies. Two parameters to model the drift velocity are the temperature and the density of the liquid, with higher velocities predicted for lower temperatures. Nonetheless, not all measurements have reported density, pressure, or purification methods. When accounting for temperature effects, the disagreement is still present, and the data points from our measurements with the purity monitor carried at 177.6 K are above those acquired at lower temperatures. Additional measurements under controlled detector conditions are necessary to shed light on the subject. For example, in previous studies it was reported that impurities diffused in the medium, from water vapour to organic molecules, can provide a more effective energy loss mechanism for electrons, with the consequence of higher mobility and decreased diffusion [39,40] (see Eqs. 9 and 10). This effect could explain the mechanism behind the unreplicated higher drift velocities in the early measurements of Guschin et. al. at 164 K [32], where no information about xenon purification methods was given. In the Xenoscope purity monitor, the long drift length and the localised source provide a measurement of the electron drift time less susceptible to non-uniformities in the electric field.

The spread of the electron cloud in time was studied by analysing the anode signal at the previously mentioned drift fields. With this purity monitor, only the longitudinal diffusion can be observed, as there is no information on the $x - y$ charge distribution. The standard deviation ($\sigma$) of the Gaussian fit of the signal in the anode is used to calculate the longitudinal diffusion coefficient. For the case of the initial and final distributions of the electron cloud following a Gaussian distribution, the width of the anode signal is related to the longitudinal diffusion [33,41,42] as:

$$D_L = \frac{d^2 \sigma_L^2}{2t^3} , \quad (11)$$

where $D_L$ is the diffusion coefficient of the electron population due to the random walk of the electrons in the longitudinal direction, and:

$$\sigma_L^2 = \sigma^2 - \sigma_0^2 , \quad (12)$$

is the width at one $\sigma$ considering diffusion effects only, which is the value to extract. The widths $\sigma_0$ and $\sigma$ belong to the initial signal at the cathode and the final signal measured in the anode, respectively, and $d$ is the drift length ($d_2 + d_3$) at drift time $t$ ($t_2 + t_3$). By rewriting Eq. (12), we obtain the width of the anode signal:

$$\sigma^2 = \frac{D_L 2t^3}{d^2} + \sigma_0^2 , \quad (13)$$

The diffusion of the electron cloud is related to the previously introduced electron mobility. At low drift fields, it follows the Einstein–Smoluchowski relation [30,43,44]:

$$D_L = \frac{k_B T}{e} \mu = \frac{\epsilon_T}{e} \mu , \quad (14)$$

with thermal energy $\epsilon_T$. The diffusion is thus affected by the cross section of elastic and inelastic interactions with the medium species (xenon and impurities), analogously to the drift velocity.

Since the electrons have energies above the thermal bath and are not in equilibrium, a characteristic energy, $\epsilon_k$, is defined as:

$$\epsilon_k = \frac{eD_L}{\mu} , \quad (15)$$

representing the energy associated with the longitudinal diffusion, where now $D_L$ has a contribution beyond the thermal energy of electrons:

$$D_L = \frac{(\epsilon_T + \epsilon)}{e}\mu , \quad (16)$$

with $\epsilon = \epsilon_k - \epsilon_T$.

Additional effects can play a role in the detected spread of the charge distribution in our detector and must be corrected to obtain a longitudinal diffusion coefficient that is independent of energy, electron source or detector response. From the original number of extracted photoelectrons to the measured signal in the anode, the following effects can impact the width:

- Duration of the pulse of the lamp, which introduces an initial signal width at one sigma of $(2.4 \pm 0.2)\,\mu s$. In this work, the initial signal width was derived from the data acquired in vacuum and in LXe, and the signal in the cathode is deconvolved with the detector and electronics responses.
- Initial asymmetry in the charge distribution given by the lamp's pulse. The effect in the width of the signal is between 1 and 10%, with higher deviations in the cathode. After the charges have drifted 50 cm, diffusion and Coulomb force between electrons contribute to a Gaussian distribution in the anode. The effect on the determination of the centre of the charge distribution for the times $t_1$ and $t_3$ affects negligibly ($\mathcal{O}(0.1\%)$) the total drift time.
- Detection response of the screening region [45]. The weighting potential between the solid electrodes and the hexagonal meshes is taken into account in the measured signal to yield the original electron cloud spread in the $z$-direction. The effect of the hexagonal screening meshes was simulated by modelling the 3D geometry of the meshes and detector and performing electrostatic simulations with COMSOL. The method to derive the weighting potential is adopted from Ref. [46]. The weighting potential is obtained by averaging the potential over different electron drift paths to smooth out local effects.
- Coulomb repulsion between electrons, where each electron is affected by the electric field induced by other electrons, can increase the size of the electron cloud. The Coulomb repulsion calculation follows the empirical approach in Ref. [47], which considers the ellipsoid explosion model from Ref. [48], where a set of differential equations is solved to obtain the final width of the signal after the drift. The repulsive forces are stronger immediately after the charge creation and become smaller as the electron cloud spreads along the drift path. After the electrons have drifted and reached the anode grid, an additional width value of 5% compared to the no repulsion forces case is inferred from the empirical approach, and taken as an uncertainty in the charge distribution after their extraction from the photocathode.
- Electron attachment to electronegative impurities which can potentially change the distribution of the electron cloud. This could in principle affect the diffusion of electrons in LXe, and the method to estimate this impact is taken from Ref. [41]. The longitudinal diffusion coefficient and drift velocity formulae are expanded to include higher-order terms containing the attachment rate. The effects of electron attachment in the diffusion can be neglected according to $\frac{D_L}{(v_d \cdot d)} \ll 1$, which is the case for this study.
- Readout response times of the pre-amplifiers. The circuit response was estimated when benchmarking the electronics and has a negligible effect on the signal shape.

Combining all the effects introduced above, a response function is obtained to deconvolve the observed signal. Table 2 compiles the systematic effects treatment for the calculation of the longitudinal diffusion coefficient. The results of this deconvolution, for each measurement at different drift fields, are shown in Fig. 9, together with literature values [47,49]. The longitudinal diffusion coefficient was measured at relatively low drift fields (i.e. $< 100$ V/cm). By using the values derived for the mobility for cold electrons and hot electrons included in Fig. 8, of $0.29 \, \text{mm}^2/(\mu\text{s V})$ and $0.01 \, \text{mm}^2/(\mu\text{s V})$, respectively, together with the thermal energy of electrons, reference coefficients for the diffusion can be obtained from Eq. (15). These are included in Fig. 9. By comparing the experimental values with the benchmark equations, the difference between these is contained in terms of the characteristic energy of electrons and their mobility, given Eq. (16).

From the discussed sources of uncertainty for the diffusion coefficient $D_L$, the largest impact originates (in terms of the variance of each parameter in the uncertainty propagation over the total variance, from 25 to 75 V/cm, respectively) from the final width measured at the anode (95–80%), the initial signal width (20–50%), the Coulomb repulsion (1–5%), the uncertainty in the drift distance (20–5%) and the drift time (4–1%).

NEST version 2.3.7 implemented an empirical model based on previous measurements for the longitudinal diffusion coefficient prediction. At the time of this analysis, NEST lacked data at low drift fields (below $\sim 100$ V/cm), and the model predicted a considerably lower longitudinal diffusion coefficient with lower drift fields, in conflict with the measured values in this and other works. An alternative model is included in NEST, based on differential cross sections derived from Dirac–Fock solutions, combined with Maxwell–Boltzmann distributions [31], which do not predict the existing experimental values at all drift fields. The subsequent NEST version 2.3.8 includes a correction on the diffusion modelling, also shown in Fig. 9, resulting from an exchange with the developers. Our analysis aimed to not only infer the values of longitudinal diffusion coefficient at low drift fields for liquid xenon, but also to understand their origin, related to the temperature and purity of the xenon.

**Table 2** Systematic effects and impact on the uncertainty for the inferred longitudinal diffusion coefficient $D_L$

| Systematic effect | Treatment and uncertainty |
| --- | --- |
| **Measured** | |
| Anode signal width, $\sigma$ | Gaussian plus sine fit, 2–3% |
| Drift time, $t_2 + t_3$ | Time interval between extrema of the cathode and anode signal fits |
| Initial signal width | Introduced by the lamp pulse. Measurements in vacuum and in LXe, $(2.4 \pm 0.2) \, \mu\text{s}$ |
| Electronics | RC time constant calculation from a square pulse, $0.2 \, \mu\text{s}$ |
| Drift length, $d_2 + d_3$ | Drift distance of the electron cloud, taken as $(513 \pm 7)$ mm when accounting for the potential contraction of the components at 177 K with an assumed 1% thermal contraction, and the position of the centre of the cloud distribution in drift regions with respect the extrema of the signals |
| Filtering and processing | Maximum 4% of anode signal width |
| Signal asymmetry | Asymmetry in the pulse from the lamp, uncertainty in the range 1-10% in the width of the signals for drift fields in the range $75 - 25$ V/cm |
| **Simulated** | |
| Detector response | COMSOL 3D model of the detector to derive the weighting potential, 10% uncertainty in the response |
| **Assumed** | |
| Coulomb repulsion | Calculated with empirical model from [47], assumption of additional 5% uncertainty in the initial signal spread |
| Electron attachment | Neglected, 4th-order correction: $\frac{D_L}{(v_d \cdot d)} \ll 1$ |

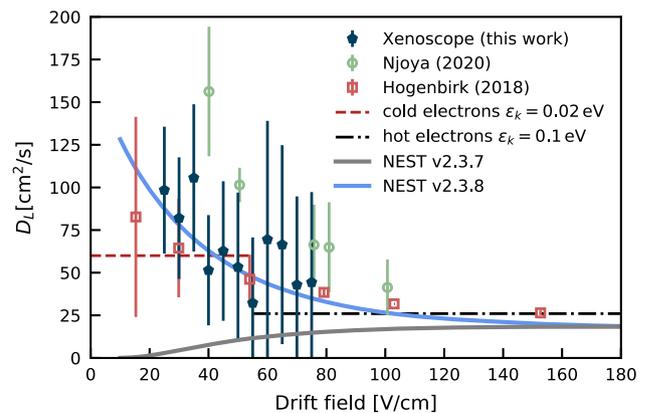

**Fig. 9** Longitudinal diffusion coefficient calculated in this work, with an electron drift lifetime of $\tau = (649 \pm 23) \, \mu\text{s}$ compared to the results from a purity monitor by Njoya et al. ($\tau \sim 1 - 35 \, \mu\text{s}$) [47], and a TPC from Hogenbirk ($\tau \sim 430 \, \mu\text{s}$) [49] and NEST [38]. The model by NEST version 2.3.7 (solid grey curve) predicted diffusion values approaching zero for lower drift fields. In version 2.3.8, a fix for this behaviour was introduced, as shown in the solid blue curve

## 5 Conclusions and outlook

The construction of a next-generation liquid xenon detector at the 50-tonne scale and beyond will face several technological challenges. To address some of these, the Xenoscope facility was designed and built to house a 2.6 m tall two-phase TPC at the University of Zurich, with a total LXe mass of $\sim 400$ kg. After a commissioning run described in Ref. [13], we presented here the first results from a run with a 53 cm tall purity monitor.

The electron drift lifetime was monitored for 88 days with varying xenon recirculation speeds. For a speed of 40 slpm, the highest achieved lifetime was $(664 \pm 23)\,\mu s$. A parametric model of the effect of the purification rate, the time-dependent outgassing rate, the liquid–gas impurity diffusion, and the injection of impurities due to operational changes, was fitted to the data. The resulting model was used to predict the electron drift lifetime evolution for different purification conditions and, therefore, inform future design and operation choices.

The electron drift velocity and the longitudinal diffusion coefficient of the electron cloud in liquid xenon were calculated based on data acquired at drift fields between 25 and 75 V/cm. With the increasing size of LXe TPCs, diffusion strongly affects the position reconstruction of events and the ability to discriminate between single and multiple interactions. Thus, accurate measurements of drift and diffusion properties, combined with an improved understanding of the systematic effect of impurity concentrations on these properties on large scales, are crucial. Our results are in agreement with previous studies both for drift velocity [32–35] and longitudinal diffusion [47,49]. They also triggered an update of NEST [38], a simulation package largely used in the community, regarding the modelling of longitudinal diffusion of electron clouds in LXe.

For the next stage of the Xenoscope project, a two-phase xenon TPC was recently built and installed, and will be operated to observe electron drift over distances up to 2.6 m. The upgrade includes liquid-level control and monitoring, high-voltage supply up to 50 kV via a commercial feedthrough, and an array of silicon photomultipliers for light-readout in the gas phase located just above the gas/liquid interface [50]. The latter replaces the charge readout used at the anode of the purity monitor, detecting instead the proportional scintillation produced in the xenon gas region of the TPC. The TPC equipped with the SiPM array will be used to study electron cloud diffusion in both longitudinal and transverse directions. Transverse diffusion is another critical parameter for more accurate modelling of electron transport in xenon-based detectors, for which measurements in the literature are scarce [33,51]. Another goal of the upgrade is to study optical properties of liquid xenon at large scales, as well as new types of photosensors under the operating conditions of DARWIN. In addition, the facility will be available to the collaboration for various R&D projects related to the realisation of a large-scale xenon TPC.

**Acknowledgements** This work was supported by the European Research Council (ERC) under the European Union's Horizon 2020 research and innovation programme, grant agreement No. 742789 (*Xenoscope*), by the SNF grant 20FL20-201437, as well as by the European Union's Horizon 2020 research and innovation programme under the Marie Skłodowska -Curie grant agreement No 860881-HIDDeN. We thank the electronics and mechanical workshops in the UZH Physics Department for their continuous support. We thank Laura Manenti for insightful discussions about purity monitors.

**Data Availability Statement** This manuscript has associated data in a data repository. [Authors' comment: The repository for the associated data is located at https://doi.org/10.5281/zenodo.8171063.]

## Appendix A: Cryostat pre-cooler

The inner vessel of the cryostat requires an initial cooling period prior to the filling of LXe. A liquid-nitrogen ($LN_2$) based pre-cooling system made of four stainless steel coolers was designed to reduce the initial cool-down time and to speed up the filling process by increasing the cooling power available to liquefy the xenon. The coolers are made of two machined stainless steel layers each, welded at their perimeter. Parallel channels are machined inside the plates

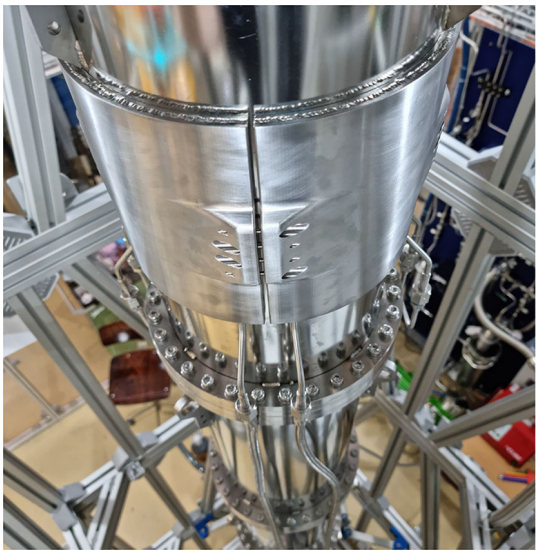

**Fig. 10** Pre-cooler installed at the top of the inner cryostat vessel. $LN_2$ is flushed from the bottom of the outer vessel through a DN40CF liquid feedthrough up to the four-quarter circle plates, connected in series. The boil-off of the $LN_2$ inside the plates provides up to $(0.57 \pm 0.03)$ kW of cooling power

to help distribute the $LN_2$ over the whole surface. Shown in Fig. 10 are four segments connected in series and attached to one another around the highest straight section of the inner vessel, making direct contact with the outside wall of the inner cryostat vessel. The assembly is compressed on the outside of the inner vessel, applying a concentric force to the cooler to ensure optimal thermal contact.

Apiezon® N Grease [52], a cryogenic thermal paste, was uniformly applied between the pre-cooler plates and the inner vessel of the cryostat to maximise the heat transfer, avoiding the trapping of air pockets which could produce virtual leaks [53]. Two 1/4" braided stainless steel hoses connect the coolers to a DN40CF liquid feedthrough located at the bottom of the outer vessel where they exit the vacuum. $LN_2$ is flushed into the coolers from a self-pressurising storage Dewar. The $LN_2$ is pushed upwards and boils off in the cooling plates, exiting in gas form by the outlet line terminated by a brass phase separator.

The maximum cooling power of the pre-cooler was calculated from the maximum achievable xenon filling speed for the purity monitor run. At all times during the fill, gaseous xenon is recirculated to and from the cryostat with a gas compressor between 25 and 40 slpm to ensure sufficient gas convection inside the inner vessel, helping the uniform cooldown of the pressure vessel. The filling speed is measured as the difference between the readings of the mass flow meter and the mass flow controller, respectively downstream and upstream of the gaseous xenon inlet [13]. The measured mass flow difference solely using the full cooling power of the PTR was measured to be $\Delta \dot{m}_{\mathrm{fill_{PTR}}} = (17.2 \pm 1.5)$ slpm, while the mass flow using both the PTR and the pre-cooler was $\Delta \dot{m}_{\mathrm{fill_{PTR+PC}}} = (70.9 \pm 1.5)$ slpm. Therefore, the increase in filling speed from the addition of the pre-cooler is $\Delta \dot{m}_{\mathrm{fill_{PC}}} = (53.7 \pm 2.2)$ slpm, neglecting any difference in the cooling rate of the inner vessel between the two filling methods as the temperature of the inner vessel is assumed constant.

The cooling power of the pre-cooler can be defined as the sum of the power required to cool the xenon down to the liquefaction point and the power needed to liquefy it:

$$P_{\mathrm{PC}} = \Delta \dot{m}_{\mathrm{fill_{PC}}} \cdot \rho_{\mathrm{GXe}} \cdot \left(C_{\mathrm{p}} \cdot \Delta T + \Delta H_{\mathrm{vap}}\right) \quad (\mathrm{A.1})$$
$$= (0.57 \pm 0.03)\,\mathrm{kW}, \quad (\mathrm{A.2})$$

where $\rho_{\mathrm{GXe}} = 5.4885$ g L$^{-1}$ is the density of GXe at 1 bar, $C_{\mathrm{p}} = 0.16067$ J g$^{-1}$ K$^{-1}$ is the heat capacity of GXe, $\Delta T = 119$ K is the temperature difference between room temperature and LXe temperature and $\Delta H_{\mathrm{vap}} = 95.587$ J g$^{-1}$ is its latent heat of vaporisation [54].

With the addition of the pre-cooler, we completed the filling of 343 kg of xenon in 20.5 h, discontinuously over a period of three days, for an average filling speed of 16.77 kg/h. With the average filling speed during the commissioning run of Xenoscope, reported in Ref. [13], estimated at 3.95 kg/h, we estimate that the addition of pre-cooler reduces the filling time by a factor 4.25.

**Appendix B: Ball of xenon**

The xenon recovery and storage system consists of two separate units: a gas recuperation system exploiting cryogenic pumping comprised of a gas bottle array and a newly-developed gravity-assisted liquid recuperation system. The former consists of an array of up to ten 40 L aluminium gas cylinders of which four hang inside interconnected Dewar flasks that can be filled with $LN_2$. The vacuum created inside the cooled cylinders allows for the recovery of the xenon from the cryostat in gaseous form. The array can store a maximum of 470 kg of xenon.

The speed of the recovery process with the gas recuperation system is limited by the evaporation rate of LXe inside the cryostat. This motivated the installation of a second unit, a liquid phase recuperation system, which allows for efficient xenon recovery and shorter downtimes between runs of Xenoscope [35]. Its main component is the spherical pressure vessel Box of Xenon (BoX) which was designed and manufactured by *KASAG Swiss AG* [55] compliant with the European Pressure Equipment Directive PED 2014/68/EU. The stainless steel sphere has a wall thickness of 15 mm and an inner radius of 450 mm. With a maximum allowable pressure of 90 bar, BoX can store up to 400 kg of xenon at room temperature. $LN_2$ can be flushed through a copper cooling block located beneath the spherical vessel to cool it down

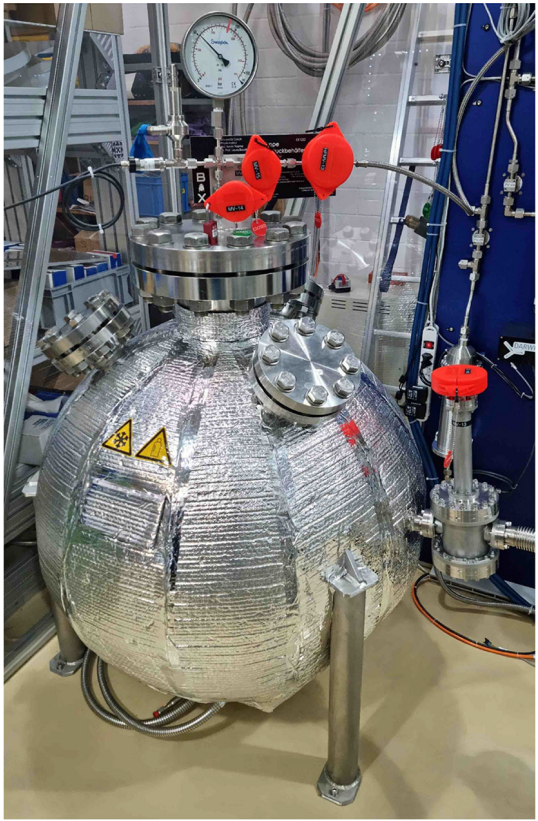

**Fig. 11** Liquid recuperation and storage system Ball of Xenon (BoX). The inner vessel of Xenoscope is isolated from BoX by two high-pressure valves, rated for cryogenic operation. During liquid recovery, LXe flows through a vacuum-insulated transfer line, visible on the right side, to the high-pressure vessel. At its top, BoX is connected to the high-pressure side of the gas system, allowing the boil-off gas to return to the cryostat. The pressure vessel is equipped with an over-pressure safety relief valve, an analogue pressure gauge and a pressure transducer which allows for pressure monitoring via the slow control system. The $LN_2$ cooler system is attached to the underside of the sphere. The vessel is thermally insulated with aluminised bubble wrap

before liquid recuperation is performed. Good thermal contact between the sphere and the copper cooler is ensured by a layer of Apiezon® N grease [52] mixed with 0.5–1.0 µm silver powder. BoX is thermally insulated with multiple layers of double-sided aluminised bubble wrap.

BoX is connected between the bottom of the inner vessel and the high-pressure side of the gas system. Unlike the gas cylinder array, BoX can thus be used to drain xenon in liquid form directly from the cryostat, assisted by gravity.

At room temperature, BoX would have a pressure of 63 bar if filled with 450 kg of xenon. Past its critical point (above 289.7 K and 58.4 bar [56]), xenon becomes supercritical and the pressure increases rapidly with temperature. The maximum allowable pressure of 90 bar will however not be exceeded below 40°C. BoX is instrumented with both an analogue pressure gauge and an electronic pressure transducer. Pressure values from the latter are sent to the slow control system for monitoring. An overpressure-relief valve to air with a set pressure of 89 bar relative to atmospheric pressure is installed as an ultimate safety system.

Prior to liquid recuperation, BoX must be pre-cooled to enhance the performance of the process. $LN_2$ is therefore flushed for $\mathcal{O}(24h)$ through the copper cooler. With the pressure in BoX at vacuum level, the recuperation process is started by opening the two valves connecting BoX to the cryostat. A fraction of the LXe coming from the cryostat evaporates after it first contacts the walls of the pressure vessel. This gas is returned to the top of the inner vessel of the cryostat to equalise the pressures of BoX and the cryostat, the gas flow is measured with the mass flow meter. As the process goes on, the evaporation rate stabilises, indicating the halting of recuperation. Xenon remaining in the inner vessel is finally cryo-pumped into the gas cylinder array.

BoX was used successfully to perform gravity-assisted LXe recuperation at the end of three operational runs. With 62.3 bar of xenon in BoX at 295.5 K after the recuperation of the purity monitor run, we estimate that approximately 294 kg of LXe were recuperated into BoX in ∼ 11.5 h, corresponding to an average recovery speed of 25.4 kg/h. This is an increase in recovery speed by factor 8 compared to gas recuperation. The addition of a liquid-height measuring device for future runs will enable the live monitoring of the liquid level during the gravity-assisted recuperation.

## References

1. DARWIN collaboration, L. Baudis, DARWIN: dark matter WIMP search with noble liquids J. Phys. Conf. Ser. **375**, 012028 (2012). https://doi.org/10.1088/1742-6596/375/1/012028. arXiv:1201.2402
2. DARWIN collaboration, J. Aalbers et al. DARWIN: towards the ultimate dark matter detector, **11**, 017 (2016). https://doi.org/10.1088/1475-7516/2016/11/017JCAP. arXiv:1606.07001
3. M. Schumann et al. Dark matter sensitivity of multi-ton liquid xenon detectors **1510**, 016 (2015). https://doi.org/10.1088/1475-7516/2015/10/016JCAP. arXiv:1506.08309
4. C.A.J. O'Hare, New definition of the neutrino floor for direct dark matter searches. Phys. Rev. Lett. **127**, 251802 (2021). https://doi.org/10.1103/PhysRevLett.127.251802. arXiv:2109.03116
5. L. Baudis, WIMP dark matter direct-detection searches in noble gases. Phys. Dark Univ. **4**, 50–59 (2014). https://doi.org/10.1016/j.dark.2014.07.001. arXiv:1408.4371
6. DARWIN collaboration, F. Agostini et al. Sensitivity of the DARWIN observatory to the neutrinoless double beta decay of $^{136}$Xe. Eur. Phys. J. C **80**, 808 (2020). https://doi.org/10.1140/epjc/s10052-020-8196-z. arXiv:2003.13407
7. R.F. Lang, C. McCabe, S. Reichard, M. Selvi, I. Tamborra, Supernova neutrino physics with xenon dark matter detectors: A timely perspective. Phys. Rev. D **94**, 103009 (2016). https://doi.org/10.1103/PhysRevD.94.103009. arXiv:1606.09243
8. N. Raj, Neutrinos from Type Ia and failed core-collapse supernovae at dark matter detectors. Phys. Rev. Lett. **124**, 141802 (2020). https://doi.org/10.1103/PhysRevLett.124.141802. arXiv:1907.05533

9. DARWIN collaboration, J. Aalbers et al. Solar neutrino detection sensitivity in DARWIN via electron scattering. Eur. Phys. J. C **80**, 1133 (2020). https://doi.org/10.1140/epjc/s10052-020-08602-7. arXiv:2006.03114
10. LZ collaboration, D.S. Akerib et al. Projected sensitivities of the LUX-ZEPLIN experiment to new physics via low-energy electron recoils. Phys. Rev. D **104**, 092009 (2021). https://doi.org/10.1103/PhysRevD.104.092009. arXiv:2102.11740
11. PandaX collaboration, H. Zhang et al. Dark matter direct search sensitivity of the PandaX-4T experiment. Sci. China Phys. Mech. Astron. **62**, 31011 (2019). https://doi.org/10.1007/s11433-018-9259-0. arXiv:1806.02229
12. XENON collaboration, E. Aprile et al. Projected WIMP sensitivity of the XENONnT dark matter experiment. **11**, 031 (2020). https://doi.org/10.1088/1475-7516/2020/11/031JCAP. arXiv:2007.08796
13. L. Baudis, Y. Biondi, M. Galloway, F. Girard, A. Manfredini, N. McFadden et al. Design and construction of Xenoscope—a full-scale vertical demonstrator for the DARWIN observatory **16**, P08052 (2021). https://doi.org/10.1088/1748-0221/16/08/P08052JINST. arXiv:2105.13829
14. C. Hasterok, Gas purity analytics, calibration studies, and background predictions towards the first results of XENON1T. PhD thesis, U. Heidelberg (2017). https://doi.org/10.11588/heidok.00023693
15. Y. Biondi, Sensitivity of DARWIN to rare events and the purity monitor for xenoscope. PhD thesis, University of Zurich (2022)
16. L7685 Xenon Lamp, Hamamatsu. https://www.hamamatsu.com/content/dam/hamamatsu-photonics/sites/documents/99_SALES_LIBRARY/etd/L7684_L6604_TLSX1029E.pdf. Datasheet, accessed July (2022)
17. LewVac Optic Fibres. https://www.lewvac.co.uk/product/fibre-optic-cable-assembly-components-accessories/. Datasheet, accessed July (2022)
18. A. Valentini, E. Nappi, M.A. Nitti, Influence of the substrate reflectance on the quantum efficiency of thin CsI photocathodes. Nucl. Instrum. Meth. A **482**, 238–243 (2002). https://doi.org/10.1016/S0168-9002(01)01678-3
19. Coater 150T Plus, Quorum. https://www.quorumtech.com/wp-content/uploads/2020/08/Q150GB_Brochure.pdf. Datasheet, accessed July (2022)
20. L. Manenti, L. Cremonesi, F. Arneodo, A. Basharina-Freshville, M. Campanelli, A. Holin et al. Performance of different photocathode materials in a liquid argon purity monitor. **15**, P09003 (2020). https://doi.org/10.1088/1748-0221/15/09/P09003JINST. arXiv:2005.08187
21. Y. Biondi, Purity monitor and TPC design for Xenoscope. Phys. Conf. Ser. **2374**, 012025 (2022). https://doi.org/10.1088/1742-6596/2374/1/012025J
22. AD8066 OPA, Analog Devices. https://www.analog.com/en/products/ad8066.html#product-overviewDatasheet, accessed July (2022)
23. Waverunner9000 Teledyne, LeCroy. https://teledynelecroy.com/oscilloscope/oscilloscopemodel.aspx?modelid=11556 Datasheet. accessed July (2022)
24. CAEN. www.https://www.caen.it/. Website, (accessed July 06, 2021)
25. COMSOL Inc. www.comsol.com. Website, accessed July (2022)
26. XENON collaboration, E. Aprile et al. Emission of single and few electrons in XENON1T and limits on light dark matter. Phys. Rev. D **106**, 022001 (2022). https://doi.org/10.1103/PhysRevD.106.022001. arXiv:2112.12116
27. LUX collaboration, D.S. Akerib et al. Investigation of background electron emission in the LUX detector. Phys. Rev. D **102**, 092004 (2020). https://doi.org/10.1103/PhysRevD.102.092004. arXiv:2004.07791
28. Z. Greene, The XENON1T Spin-independent WIMP dark matter search results and a model to characterize the reduction of electronegative impurities in its 3.2 tonne liquid xenon detector. PhD thesis, Columbia University, (2018). https://doi.org/10.7916/D87M1RTN
29. G. Plante, E. Aprile, J. Howlett, Y. Zhang, Liquid-phase purification for multi-tonne xenon detectors. Eur. Phys. J. C **82**, 860 (2022). https://doi.org/10.1140/epjc/s10052-022-10832-w. arXiv:2205.07336
30. W.F. Schmidt, Electronic Conduction Processes in Dielectric Liquids. IEEE. Trans. Electric. Insul. **19**, 389–418 (1984). https://doi.org/10.1109/TEI.1984.298767
31. G. Boyle, R. McEachran, D. Cocks, M. Brunger, S. Buckman, S. Dujko et al., Ab-initio electron scattering cross-sections and transport in liquid xenon. Phys. D **49**, 5201 (2016). https://doi.org/10.1088/0022-3727/49/35/355201J. arXiv:1603.04157
32. E.M. Gushchin, A.A. Kruglov, I.M. Obodovskii, Electron dynamics in condensed argon and xenon. Sov. Phys. JETP **55**(55), 650 (1982)
33. EXO- 200 collaboration, J. B. Albert et al. Measurement of the drift velocity and transverse diffusion of electrons in liquid xenon with the EXO-200 detector. Phys. Rev. C **95**, 025502 (2017). https://doi.org/10.1103/PhysRevC.95.025502. arXiv:1609.04467
34. L. Baudis et al., A Dual-phase Xenon TPC for Scintillation and Ionisation Yield Measurements in Liquid Xenon. Eur. Phys. J. C **78**, 351 (2018). https://doi.org/10.1140/epjc/s10052-018-5801-5. arXiv:1712.08607
35. K. Thieme, The Low-Energy and Large-Scale Frontier of Dual-Phase Xenon Time Projection Chambers for Dark Matter Search. PhD thesis, University of Zurich, (2022)
36. F. Jörg, D. Cichon, G. Eurin, L. Hötzsch, T. Undagoitia Marrodán, N. Rupp, Characterization of alpha and beta interactions in liquid xenon. Eur. Phys. J. C **82**, 361 (2022). https://doi.org/10.1140/epjc/s10052-022-10259-3. arXiv:2109.13735
37. D. Baur et al. The XeBRA platform for liquid xenon time projection chamber development. **18**, T02004 (2023). https://doi.org/10.1088/1748-0221/18/02/T02004JINST. arXiv:2208.14815
38. M. Szydagis, J. Balajthy, J. Brodsky, J. Cutter, J. Huang, E. Kozlova et al. Noble Element simulation technique v2.0. https://doi.org/10.5281/zenodo.1314669
39. K. Yoshino, U. Sowada, W.F. Schmidt, Effect of molecular solutes on the electron drift velocity in liquid Ar, Kr, and Xe. Phys. Rev. A **14**, 438–444 (1976). https://doi.org/10.1103/PhysRevA.14.438
40. J.L. Pack, A.V. Phelps, Drift Velocities of Slow Electrons in Helium. Neon, Argon, Hydrogen, and Nitrogen, Phys. Rev. **121**, 798–806 (1961). https://doi.org/10.1103/PhysRev.121.798
41. Y. Li et al. Measurement of Longitudinal Electron Diffusion in Liquid Argon. Nucl. Instrum. Meth. A **816**, 160–170 (2016). https://doi.org/10.1016/j.nima.2016.01.094. arXiv:1508.07059
42. W. Blum, L. Rolandi, W. Riegler, Particle detection with drift chambers. Particle Acceleration and Detection. Springer Berlin Heidelberg, (2008). https://doi.org/10.1007/978-3-540-76684-1
43. A. Einstein, Über die von der molekularkinetischen Theorie der Wärme geforderte Bewegung von in ruhenden Flüssigkeiten suspendierten Teilchen. Annalen der Physik **322**, 549–560 (1905). https://doi.org/10.1002/andp.19053220806
44. M. von Smoluchowski, Zur kinetischen theorie der brownschen molekularbewegung und der suspensionen. Annalen der Physik **326**, 756–780 (1906). https://doi.org/10.1002/andp.19063261405
45. W. Shockley, Currents to conductors induced by a moving point charge. Appl. Phys. **9**, 635–636 (1938). https://doi.org/10.1063/1.1710367J
46. A. Göök, F.-J. Hambsch, A. Oberstedt, S. Oberstedt, Application of the Shockley–Ramo theorem on the grid inefficiency of Frisch grid ionization chambers. Nucl. Instrum. Meth. A **664**, 289–293 (2012). https://doi.org/10.1016/j.nima.2011.10.052. arXiv:1508.07059

47. O. Njoya et al., Measurements of electron transport in liquid and gas Xenon using a laser-driven photocathode. Nucl. Instrum. Meth. A **972**, 163965 (2020). https://doi.org/10.1016/j.nima.2020.163965. arXiv:1911.11580
48. M. Grech, R. Nuter, A. Mikaberidze, P. Di Cintio, L. Gremillet, E. Lefebvre et al., Coulomb explosion of uniformly charged spheroids. Phys. Rev. E **84** (2011). https://doi.org/10.1103/PhysRevE.84.056404. arXiv:1105.0409
49. E. Hogenbirk, M. P. Decowski, K. McEwan, A. P. Colijn, Field dependence of electronic recoil signals in a dual-phase liquid xenon time projection chamber, **13**, P10031 (2018). https://doi.org/10.1088/1748-0221/13/10/P10031JINST. arXiv:1807.07121
50. R. Peres, SiPM array of Xenoscope, a full-scale DARWIN vertical demonstrator. JINST (2023). https://iopscience.iop.org/article/10.1088/1748-0221/18/03/C03027
51. E. Aprile, T. Doke, Liquid xenon detectors for particle physics and astrophysics. Rev. Mod. Phys. **82**, 2053–2097 (2010). https://doi.org/10.1103/RevModPhys.82.2053. arXiv:0910.4956
52. Apiezon N Grease. https://apiezon.com/products/vacuum-greases/apiezon-n-grease/. Website, (accessed October 23, 2022)
53. D. Edwards, The influence of virtual leaks on the pressure in high and ultra-high vacuum systems. Vacuum **29**, 169–172 (1979). https://doi.org/10.1016/S0042-207X(79)80748-4
54. Gas Encyclopedia Air Liquide. https://encyclopedia.airliquide.com/xenon. Website, (accessed October 23, 2022)
55. KASAG Swiss AG. https://www.kasag.com/. Website, (accessed 2023, February 3)
56. P. J. Linstrom, W. G. Mallard, eds., Standard Reference Database Number 69, vol. 20899. NIST, retrieved December 10, 2020, https://doi.org/10.18434/T4D303